\documentclass[preprint,onecolumn,showpacs,preprintnumbers,amsmath,amssymb]{revtex4-1}

\usepackage{subfigure}
\usepackage{epsf}
\usepackage{graphicx}

\newcommand{\be}{\begin{equation}}
\newcommand{\ee}{\end{equation}}
\newcommand{\bea}{\begin{eqnarray}}
\newcommand{\eea}{\end{eqnarray}}

\begin{document}
 
\title{Surface and bulk polaritons in a linear magnetoelectric multiferroic with canted spins: Transverse Electric polarisation}

\author{V. Gunawan}
\email{slamev01@physics.uwa.edu.au}
\author{R.~L.~Stamps}

\affiliation{School of Physics M013, University of Western Australia, 35 Stirling Hwy, Crawley WA 6009, Australia}

\date{\today}

\begin{abstract}
Some magnetoelectric multiferroics have a canted spin structure that can be described by a Dzyaloshinkii-Moriya coupling.  We calculate properties and features expected for surface and bulk magnon polaritons in such media with a linear magnetoelectric interaction for the case of transverse electric polarisation. The dielectric polarisation and magnetisation of weak ferromagnetism are constrained to lie in the plane parallel to the surface.  We examine a geometry with the polarisation oriented in the film plane and present numerical results for the transverse electric polarisation. Particular attention is given to  non-reciprocal surface modes, which exist in frequency between two bulk bands, and show how these modes can be modified by  external magnetic field. Results for attenuated total reflection are presented, and discussed in relation to nonreciprocity.  Example results are calculated for the canted antiferromagnet BaMnF4.
\end{abstract}

\pacs{71.36+c; 78.20.Jq; 78.20.Ls}

\maketitle
%-----------------------------------------INTRODUCTION-----------------------------------------------------------------------------------------%
\section{Introduction}
\label{Introduction}

Polaritons in magnetic and dielectric media arise from coupling between the electromagnetic field and polarisations of the media\cite{camley82}.  A particularly interesting polariton is localised to the surface of a bounded material.  Surface polaritons can display nonreciprocity, where the propagation frequency in one direction is different from the propagation in the opposite direction.  If the propagation wavevector is $\vec{k}$, then nonreciprocity can be expressed in terms of frequency $\omega$ as $\omega\left(\vec{k}\right)\neq\omega\left(-\vec{k}\right)$. Numerous applications for nonreciprocal surface waves exist\cite{welford91,cairns99} including recent developments  for surface polariton optics\cite{keilmann98,konopsky06}.

Magnon polaritons are coupled photon and magnetic excitations, and exist for simple ferromagnets\cite{damon60,harstein73} and multisublattice magnets including antiferromagnets\cite{abraha96}.  In the present work, we discuss surface modes for an interesting class of materials that are currently a focus of attention: magnetic multiferroics. Multiferroics display long range order and polarisable response in two or more aspects: elastic, dielectric and magnetic. The calculations presented here are illustrated using parameters appropriate for BaMnF$_4$, a material that has been modeled with a linear magnetoelectric coupling between the magnetic and dielectric subsystems.  This material has two magnetic sublattices, and a canting angle between sublattice magnetisations results from magnetoelectric interaction\cite{barnas86a,barnas86b,tilley82}.   

Some properties of bulk polaritons in linear magnetoelectric coupled media have been calculated  theoretically in Refs. \onlinecite{barnas86a,barnas86b}, and surface modes have been examined in Refs. \onlinecite{buchel86,tarasenko00}.  To the best of our knowledge, until now dispersion relations in the presence of applied electric and magnetic fields have not been discussed, and explicit results for surface modes in BaMnF$_4$ have not been presented. 

In this work, we show that nonreciprocity of magnetic polariton surface modes can be modified by application of an external magnetic field.  We extend  previous work by allowing magnetic sub-lattices to cant, and provide explicit results for surface and bulk modes on BaMnF$_4$.   Our results are relevant for understanding and predicting microwave and infrared responses of ferroelectric/magnetic and multiferroic/magnetic multilayers\cite{murugavael05,tabata98} and composites\cite{bai09,bichurin02}.   In these types of heterostructures, it may be possible to achieve effective media with magnetoelectric properties and a range of material parameter values not possible in single phase multiferroics.
	
The paper is organised as follows.  The field dependence of the magnetic canting angle is discussed in the next section, and relevant magnetic and dielectric susceptibilities are calculated in section III. Results for surface and bulk mode dispersions are presented and discussed in the final section IV.

%--------------------------------------------------------------------------------------------------------Section II Geometry--------------------------%
\section{Geometry and Magnetic sub-lattice canting}
\label{Geometry and Magnetic sub-lattice canting} 
%--------------------------------------------------------------------------------------------------------figure 1 --------------------------%

\begin{figure}[ht]
\includegraphics[width=7cm]{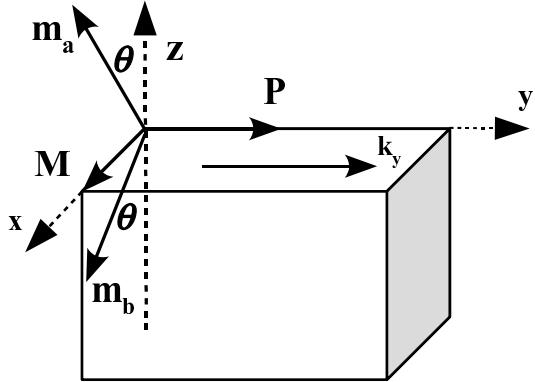}
\caption{\label{geometry}Geometry. Canting of two magnetic sublattices ($m_a$ and $m_b$) by an angle $\theta$ produces a weak ferromagnetism ($M$) along the $\hat{x}$ axis parallel to the surface. The spontaneous polarization ($\vec{P}$) is assumed to lie in planes parallel to the surface and propagation of the surface mode is along the $\hat{y}$ axis with wavenumber $\vec{k}_y$.}
\end{figure}
%----------------------------------------------------------------------------------------------------------------------------------%

The geometry used is shown in Fig. \ref{geometry}.  A semi-infinite film of a multiferroic with a two sub-lattice antiferromagnet fills the half space $z<0$ while the other half space is assumed to be vacuum.  The  magnetocrystalline easy axis for a uniaxial anisotropy is directed out of plane along the $z$ axis.  The spin system cants symmetrically in the $x-z$ plane, and the sublattices are assumed to have the same magnitude of magnetisation, $\left|\vec{m}_a\right|=\left|\vec{m}_b\right|=M_s$ .  The nett magnetisation, $\vec{M}$ corresponding to the weak ferromagnet due to spin canting, lies along the $x$ axis in a plane parallel to the surface.  The spontaneous dielectric polarisation is oriented in the perpendicular direction, also in the $x-z$ plane.  An external magnetic field is applied parallel to the weak ferromagnetic moment.

The transverse electric (TE) mode is associated with the condition where the electric part of electromagnetic waves propagate along the surface is parallel to the surface.   We assume that the surface modes propagate on the surface along the $\hat{y}$ direction, hence the electric components lie on the surface in the $\hat{x}$ direction while the magnetic part has $H_y$ and $H_z$ components.

A fourth order Landau-Ginzburg density energy is assumed for the spontaneous ferroelectric polarisation:
\begin{equation}
	F_e=\frac{1}{2}\zeta_1P^{2}_{y}+\frac{1}{4}\zeta_2P^{4}_{y}-P_yE_y.
\end{equation}
The coefficient $\zeta_1$ is temperature dependent and an external static applied electric field $E_y$ along spontaneous polarisation is included.

The magnetic contribution to the density energy is given by:
\begin{equation}
\label{fm}
F_M=-\lambda\vec{m}_a\cdot\vec{m}_b-\frac{K}{2}\left[\left(\vec{m}_a\cdot\hat{z}\right)^{2}+\left(\vec{m}_b\cdot\hat{z}\right)^{2}\right]-\left(\vec{m}_a+\vec{m}_b\right)_x H_o
\end{equation}
where $\lambda$ and $K$ represent magnetic exchange and anisotropy energies, $\vec{m}_i$ is the magnetisation of sub-lattice $i$ and $H_o$ is an external static magnetic  field. The first and second terms in the right hand side Eq. (\ref{fm}) represent the exchange and anisotropy energy, while the third term is Zeeman energy contributed by external magnetic field.  In this configuration, the easy axis is out of plane. We assume that there is zero nett magnetisation along the $\hat{z}$ direction and do not consider demagnetisation fields.  

The magneto-electric coupling is assumed to be of the  Dzyaloshinkii-Moriya (DM) form, appropriate to materials such as FeTiO$_3$\cite{ederer08} and BaMnF$_4$\cite{barnas86a,barnas86b,tilley82}. We suppose a linear coupling of the form 
{\setlength\arraycolsep{2pt}
\begin{eqnarray}
\label{ME}
&F_{ME}&=-\alpha P_{y}M_{x}L_{z}
\nonumber\\
& &=-\alpha P_{y}\cdot\left(\vec{m}_a\times\vec{m}_b\right)-\alpha P_{y}\left[\left(m_a\right)_x\left(m_a\right)_z-\left(m_b\right)_x\left(m_b\right)_z\right]
\end{eqnarray}}
where
\begin{equation}
M_x=\left(\vec{m}_a+\vec{m}_b\right)_x=2M_s\sin\theta,
\end{equation}
\begin{equation}
L_z=\left(\vec{m}_a-\vec{m}_b\right)_z=2M_s\cos\theta
\end{equation}
and $\alpha$ is the magneto-electric coupling constant.  The component  $M_x$ is the transverse component of magnetisation and represents the weak ferromagnetism.  The component $L_z$ is the magnitude of the longitudinal component of the magnetisation.   The first term in Eq. (\ref{ME}) represents the polarisation induced Dzyaloshinkii-Moriya interaction in our model, and we assume that this term is responsible for generating  canting in the magnetic sublattices.  

The canting angle is calculated by minimizing the magnetic and magnetoelectric free energies.  Writing $\frac{\partial}{\partial\theta}\left(F_M+F_{ME}\right)=0$, we obtain the condition:
\begin{equation}
\label{eqsdt}
H_o \cos\theta-\frac{1}{2}KM_s\sin2\theta+2\alpha P_yM_s\cos2\theta+\lambda M_s\sin2\theta =0.
\end{equation}
This can be written in the form
\begin{equation}
\label{sdt}
\tan2\theta=\frac{4\alpha P_y}{K-2\lambda}
\end{equation}
when the external magnetic field is absent, $H_o=0$.  As can be seen from Eq. (\ref{sdt}) above, the magnitude of the magnetoelectric coupling and spontaneous polarisation control the canting angle.

The equilbrium spontaneous polarisation is found by minimizing the dielectric and magnetoelectric free energies, $\frac{\partial}{\partial P}\left(F_E+F_{ME}\right)=0$, with respect to the spontaneous polarisation, with the result:
\begin{equation}
\label{equil}
\zeta_1 P_y+\zeta_2 P^{3}_{y}-2\alpha M^{2}_{s}\sin2\theta-E_y=0.
\end{equation}
Lastly, the temperature dependent magnetisation in mean field is calculated using the Brillouin function in $B\left(\eta\right)$
\begin{equation}
\label{bril}
M_s\left(T\right)=M_{s}\left(0\right)B_{s}\left(\eta\right)
\end{equation}
where 
\begin{equation}
\eta=\frac{g\mu_B S}{k_B T}=\left[-\lambda M_s\cos2\theta+KM_s\cos^2\theta+2\alpha P_y M_s\sin2\theta+H_o\sin\theta\right].
\end{equation}
The spontaneous polarization, magnetization and canting angle are obtained by solving Eqs. (\ref{eqsdt}), (\ref{equil}) and (\ref{bril}) simultaneously using numerical root finding techniques.

%----------------------------------------------------------------------------------------------------------------------------------%
\section{Dynamic Susceptibilities}
\label{Dynamic Susceptibilities} 
The electromagnetic problem requires constitutive relations for magnetisations and polarisations. These are formulated in terms of linear magnetic and dielectric suceptibilities and are derived from the magnetic Bloch equations:
\begin{equation}
\dot{\vec{M}}=\gamma\times\left(\frac{-\partial}{\partial\vec{M}}\left(F_M+F_{ME}\right)\right)
\end{equation}
and the Landau-Khalatnikov dynamic equations for the polarisation $P$:
\begin{equation}
\ddot{\vec{P}}=-f\frac{\partial}{\partial\vec{P}}\left(F_E+F_{ME}\right).
\end{equation}
Here $\gamma$ is the gyromagnetic ratio and $f$ is the inverse phonon effective mass.  

The relevant equations of motion for TE modes after linearisation are,
\begin{equation}
\label{dyn1}
-i\omega l_x=2\gamma M_s h_y\cos\theta-\left(\omega_a\cos\theta+2\omega_{me}\sin\theta-2\omega_{ex}\cos\theta\right)m_y,
\end{equation}
\begin{equation}
-i\omega m_y=2\gamma M_s h_z\sin\theta+\left(\omega_a\cos\theta+2\omega_{me}\sin\theta\right)l_x+\left(\omega_a\sin\theta-2\omega_{me}\sin\theta-\omega_o\right)m_z,
\end{equation}
\begin{equation}
-i\omega m_z=2\gamma M_s h_y\sin\theta+\left(\omega_o+2\omega_{me}\cos\theta\right)m_y
\end{equation}
and
\begin{equation}
\label{dyn2}
\frac{-\omega^2}{f}p_x=-\left(\zeta_1+\zeta_2P^{2}_{o}\right)p_x+e_x
\end{equation}
where $l_i=m_{a,i}-m_{b,i}$ and $m_i=m_{a,i}+m_{b,i}$.  The frequencies $\omega_a=\gamma K M_s$, $\omega_{ex}=\gamma\lambda M_s$, $\omega_{me}=\gamma\alpha P_oM_s$ and $\omega_o=\gamma H_o$ are associated with the magnetic anisotropy, exchange, magnetoelectric coupling and the external magnetic field respectively.  The parameter $P_o$ represents the equilibrium spontaneous polarisation.  It can be seen from Eqs.(\ref{dyn1}) to (\ref{dyn2}), that the dynamic magnetic and electric components are not coupled directly.  Hence, the susceptibilities can be written as
\begin{equation}
	\vec{m}=\chi^m\vec{h} \qquad \textrm{and}  \qquad \vec{p}=\chi^e\vec{e},
\end{equation}
with magnetic components given by
\begin{equation}
\label{sucepty}
	\chi _{y}^{m}=\frac{2\gamma M_{s}\left(\omega _{a}\cos 2\theta +2\omega_{me}\sin{2\theta}+\omega
_{o}\sin \theta \right)}{\left(\omega _{\mathit{afm}}^{2}\cos
^{2}\theta +\Omega _{\mathit{me}}^{2}+\Omega _{o}^{2}-\omega
^{2}\right)}
\end{equation}

\begin{equation}
	\chi _{z}^{m}=\frac{2\gamma M_{s}\left(\omega _{\mathit{me}}\sin
2\theta +\omega _{o}\sin \theta \right)}{\left(\omega
_{\mathit{afm}}^{2}\cos ^{2}\theta +\Omega _{\mathit{me}}^{2}+\Omega
_{o}^{2}-\omega ^{2}\right)}
\end{equation}

\begin{equation}
\label{suceptyz}
	\chi _{\mathit{yz}}^{m}=-\chi _{\mathit{zy}}^{m}=\frac{i2\gamma
M_{s}\left(\omega \sin \theta \right)}{\left(\omega
_{\mathit{afm}}^{2}\cos ^{2}\theta +\Omega _{\mathit{me}}^{2}+\Omega
_{o}^{2}-\omega ^{2}\right)}
\end{equation}
Here  $\omega _{\mathit{afm}}^{2}=\omega _{a}\left(\omega _{a}-2\omega
_{\mathit{ex}}\right)$ represents the antiferromagnet (AFM) resonance
frequency. The frequency  $\Omega _{\mathit{me}}$ is defined as
\begin{equation}
	\Omega _{\mathit{me}}^{2}=2\omega _{\mathit{me}}\left(2\omega
_{\mathit{me}}+\frac{1}{2}\omega _{a}\sin 2\theta -\omega
_{\mathit{ex}}\sin 2\theta \right)
\end{equation}
The frequency  $\Omega _{o}$ is the frequency shift due to an external
magnetic field and is given by
\begin{equation}
\label{Wh}
\Omega _{o}^{2}=\omega _{o}\left[\omega _{o}-\omega _{a}\sin \theta+4\omega _{\mathit{me}}\cos \theta \right]
\end{equation}
Even though there is no direct coupling between magnetic and electric parts represented by a ME susceptibility, the ME $\omega_{me}$ still appears in the magnetic susceptibility, as seen in Eq. (\ref{sucepty}) to (\ref{suceptyz}).  This results from the fact that the canting condition involves the electric polarisation.  Finally, the electric susceptibility for  $\chi _{\mathit{xx}}^{e}$ is
\begin{equation}
	\chi _{\mathit{xx}}^{e}=\frac{\omega _{L}^{2}-\omega ^{2}}{\omega
_{T}^{2}-\omega ^{2}}
\end{equation}
where $\omega _{L}$ is the frequency of the longitudinal phonon mode, and $\omega _{T}$ is the frequency of the transverse phonon mode.

%----------------SECTIONIV--------------------------------------------------------------------------------------------------%
 
\section{Results and Discussion}
\label{Results and Discussion} 
 
Since there is no dynamic ME coupling, the dispersion relation for bulk
polaritons can be calculated from the macroscopic electromagnetic wave equation
\begin{equation}
\label{wave}
\nabla ^{2}\vec{H}-\nabla \left(\nabla \cdot
\vec{H}\right)-\frac{\epsilon }{c^{2}}\vec{\mu }\cdot \partial
^{2}\vec{H}/{\partial t^{2}}=0.
\end{equation}
For the TE modes, the constitutive connecting  $\vec{B}$ with  $\vec{H}$
are defined by  $\vec{\mu }=1+4\pi \vec{\chi }_{m}$. \ Using these
relations, the solution of the wave equations for the bulk modes is of
the form
\begin{equation}
\vec{H}\sim e^{i\left(k_{y}y+k_{z}z-\omega
t\right)}
\end{equation}
and an implicit expression for the bulk mode frequency is:
\begin{equation}
\label{bulk}
\mu _{y}k_{y}^{2}=\epsilon _{x}\left(\frac{\omega}{c}\right)^{2}\left(\mu _{y}\mu _{z}-{\mu_{yz}}^{2}\right).
\end{equation}
Equation (\ref{bulk}) has two zeros, one is for  $\epsilon _{x}=0$ and the other for  $f\left(\mu \right)=\mu _{y}\mu _{z}-\mu _{\mathit{yz}}^{2}=0$. It also has two resonance poles from $\epsilon _{x}$ and another from the condition  $\mu _{y}=0$. As a result, there are three bands of
bulk polariton modes.

The dispersion relation for surface modes is obtained by assuming solutions in the form:
\begin{equation}
\label{solb}
\vec{H}\sim e^{\beta z}e^{i\left(k_{y}y-\omega t\right)}\qquad
\textrm{for $z<0$}
\end{equation}
and
\begin{equation}
\vec{H}\sim e^{-\beta _{o}z}e^{i\left(k_{y}y-\omega t\right)}\qquad
\textrm{for $z>0$}
\end{equation}
where $\beta$ and $\beta _{o}$ are positive real attenuation constants for the sample and vacuum, respectively. An implicit relation for the attenuation factor $\beta$ of the medium is derived by substituting Eq. (\ref{solb}) into the wave equation Eq. (\ref{wave}):
\begin{equation}
\label{attb}
\mu _{z}\beta ^{2}=\mu _{y}k_{y}^{2}-\epsilon _{x}\left(\frac{\omega
}{c}\right)^{2}\left(\mu _{y}\mu _{z}-\mu
_{\mathit{yz}}^{2}\right)
\end{equation}
An explicit relation for the attenuation constant
${\beta_o}$ (in vacuum) is given by
\begin{equation}
\label{attbo}
\beta _{o}^{2}=k_{y}^{2}-\left(\frac{\omega}{c}\right)^{2}.
\end{equation}
An implicit surface mode dispersion relation is calculated by requiring continuity of tangential $\vec{H}$ and normal  $\vec{B}$ at the interface $z=0$. Using Eq.(\ref{attb}) and (\ref{attbo}), we find
\begin{equation}
\label{surface}
\mu _{z}\beta +\left(\mu _{y}\mu _{z}-\mu
_{\mathit{yz}}^{2}\right)\beta _{o}+\mu_{\mathit{yz}}k_{y}=0.
\end{equation}
The dispersion implied by Eq. (\ref{surface}) involves mainly of magnetic permeabilities, and so describes magnetic surface polaritons for the weak ferromagnet. From symmetry, we know that magnetic surface polaritons can be nonreciprocal\cite{camley87}, and this is the case here as well. The direction of  $\vec{k}_{y}$ matters in Eq. (\ref{surface}) and indicates nonreciprocal propagation.

A key point is that the nonreciprocity of the surface modes depends strongly on the canting angle. If the canting angle is zero, then $\mu _{\mathit{yz}}\rightarrow 0$ and  $\mu _{z}\rightarrow 1$, and the
dispersion relation in Eq. (\ref{surface}) becomes
\begin{equation}
\beta +\mu _{y}\beta_{o}=0
\end{equation}
illustrating the dispersion of the surface modes in this case is reciprocal under reversal of  $\vec{k}$, and propagation of the surface modes is reciprocal. 

%--------------------------------------------------------------------------------------------------------figure 2 --------------------------%
\begin{figure}[ht]
\begin{center}
\subfigure[\label{attenuation}Attenuation constant.]{
\includegraphics[width=7cm]{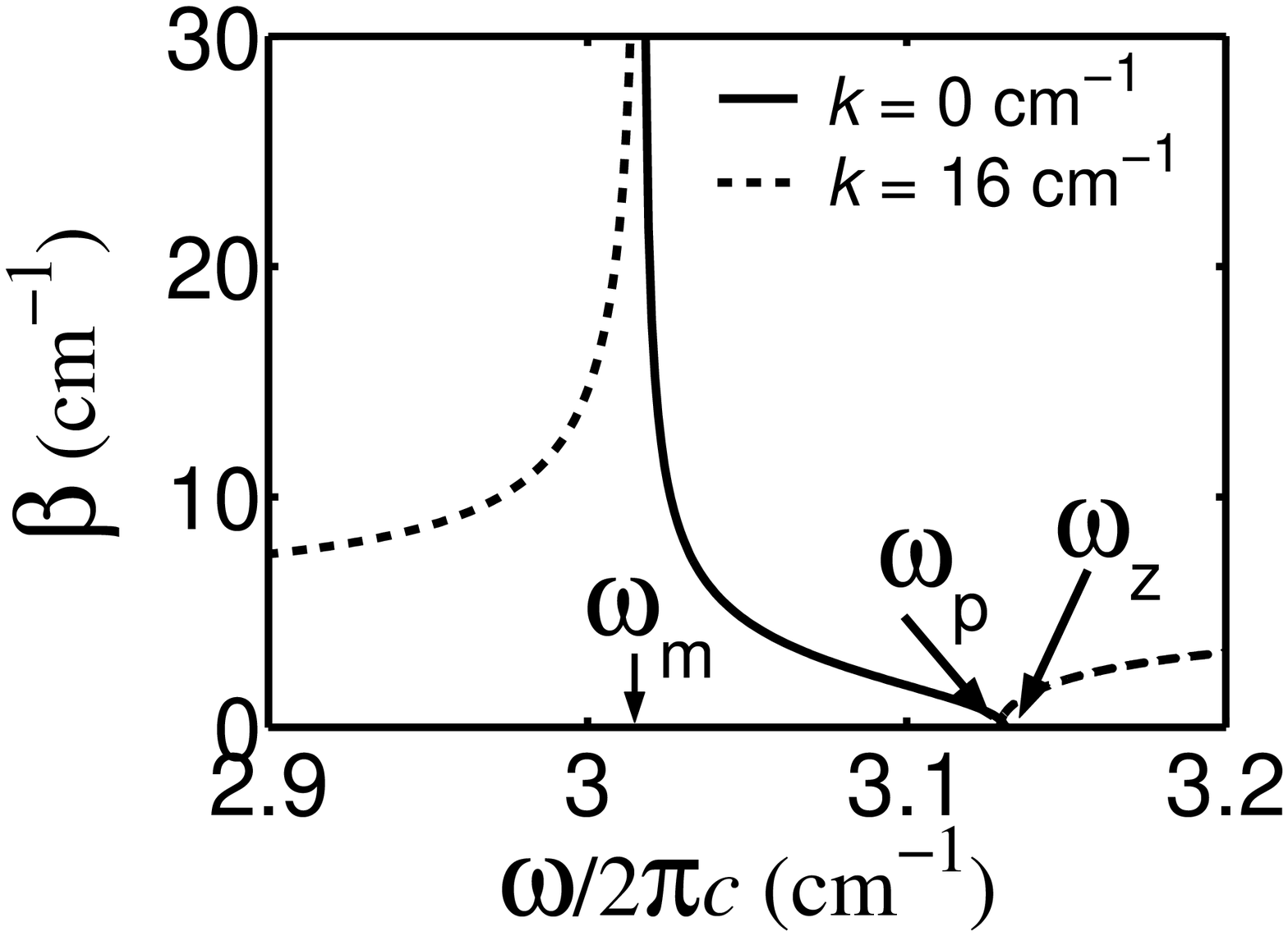}}
\subfigure[\label{dispersion}Dispersion relation.]{
\includegraphics[width=7.5cm]{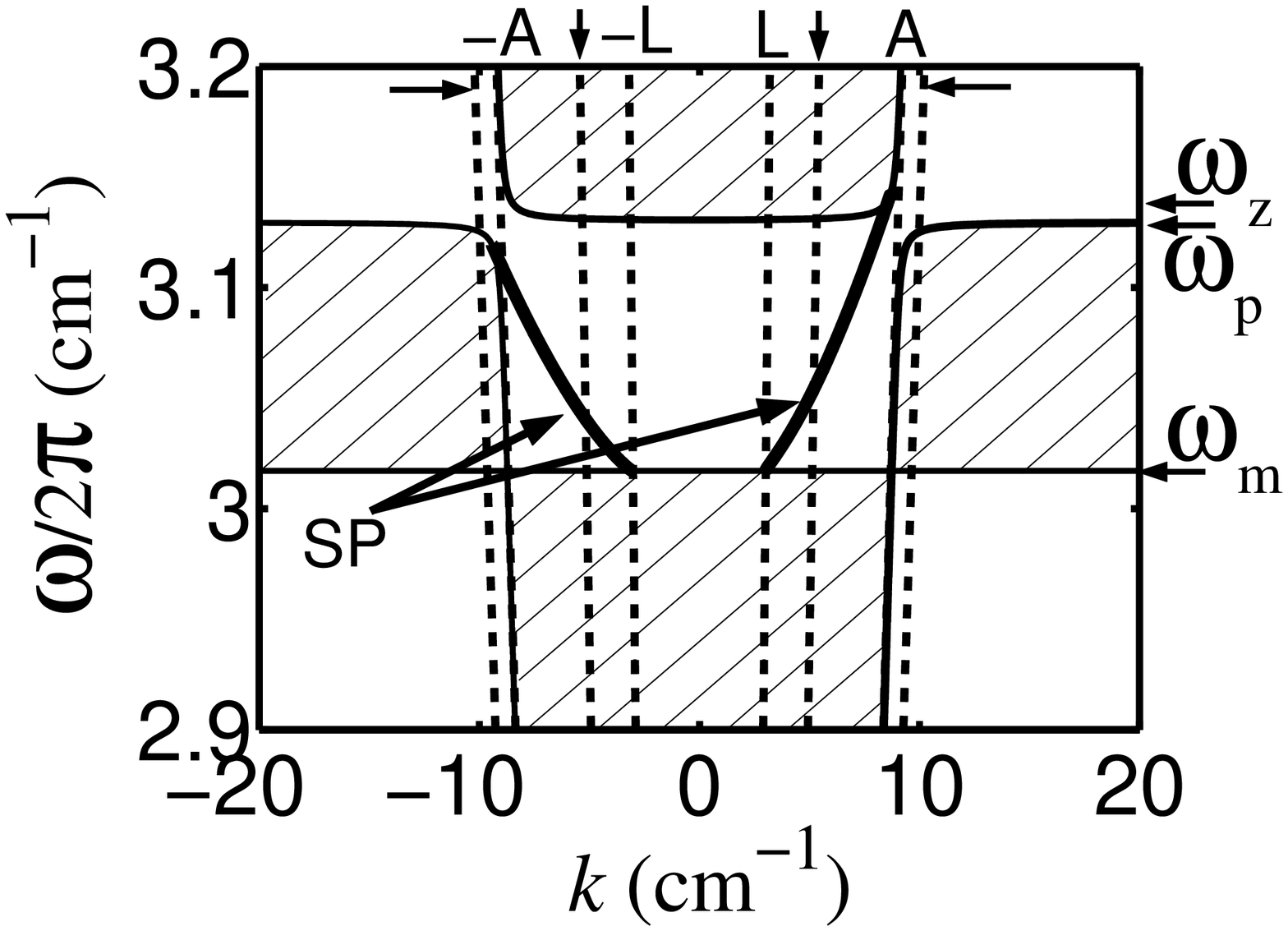}}
\caption{Attenuation constant and Dispersion relation.  In (a) the attenuation constant is shown for two values of wavevector in the absence of external fields.  The solid line represents  k = 0  cm$^{-1}$, while the dashed line correspond to  k = 16 cm$^{-1}$ .  In (b) bulk and surface mode dispersions are shown. Surface modes are indicated by "SP".  The shaded regions represent bulk bands, limited by frequencies $\omega_m$, $\omega_p$ and $\omega_z$.   The dashed lines denote by vertical and horizontal arrows represent ATR light lines for incident angles of 30$^o$ and 70$^o$.  The asymptotic boundaries and lightline are represented by dashed lines which are indicated as "A" and by "L".}
\end{center}
\end{figure}
%----------------------------------------------------------------------------------------------------------------------------------%

Solutions of Eq. (\ref{bulk}) and Eq. (\ref{surface}) are plotted in Fig. \ref{dispersion} for the case of no applied fields. Before continuing, we discuss our choice of parameters, which were chosen to be appropriate for BaMnF$_4$. The sublattice magnetisation is approximated by $M_c=2M_s\sin\theta $. Assuming the weak ferromagnetisation, as mention in review paper by Scott\cite{scott79}, $M_c=1460$ A/m with a canting angle of 3 mrad at 4.2 K as reported by Venturini\cite{venturini}, results in $M_{s}=3.054\times10^3$ Oe.  The exchange field is estimated by Holmes\cite{holmes69}, as $H_E$=50 T.  Using the definition $H_E= -\lambda M_s$, we obtain $\lambda $= -163.72.  Since antiferromagnet resonance was reported\cite{samara76} at around 3cm$^{-1}$, the anisotropy constant is estimated by using expression $\omega_r^2=\gamma^2 K(K-2\lambda)M_s^2$, with the result that $K=0.337$. Here, we estimate the gyromagnetic ratio by calculating $\gamma=g\mu_B/\hbar$, yielding $\gamma=0.933$ cm$^{-1}$T$^{-1}$.

The spontaneous ME coupling  $\alpha $ is calculated from Eq. (\ref{sdt}) using the calculated spontaneous polarisation\cite{keve71}, $P_{o}$=0.115 C/m\textsuperscript{2}, from which we obtain $\alpha $= 4.27 m\textsuperscript{2}/C.  The phonon inverse mass is calculated using the relation between phonon frequency in transversal and longitudinal modes, $\omega^{2}_{L}=\omega^{2}_{T}+4\pi f$, yielding $f=$6.4458 cm$^{-2}$ (where $\omega_{L}= 41$ cm$^{-1}$ and $\omega_{T}= 40$ cm$^{-1}$ as given in Ref. [\onlinecite{samara76}] and [\onlinecite{scott79}]).  The Ginzburg-Landau constants $\zeta _{1}$ and  $\zeta _{2}$ are calculated by solving Eq. (\ref{equil}) with the equation for the soft phonon frequency,  $\left(\zeta _{1}+3\zeta_{2}P_{o}^{2}\right)f=\omega _{L}^{y}$, resulting in $\zeta _{1}= -1.22\times10^{2}$ and $\zeta_{2}=1.085\times10^{-7}$ cm$^4$/statC$^2$.

We also note that previous investigations\cite{samara76} have assumed the dielectric constant in the  $\hat{x}$ direction to be independent of frequency with the value  $\epsilon_x=8.3$. This assumption is valid if the dielectric and magnetic responses lie in very different frequency ranges. As a result, only two bands for bulk polaritons can exist since only one pole and one zero are contributed by the magnetic sub-system. We make this assumption in what follows, since BaMnF$_4$ has a wide separation in frequency between the dielectric and magnetic resonances.  At the end of this section we discuss the consequences of relaxing this assumption.

The bulk mode frequencies are affected by the attenuation constant, since this parameter determines the frequency and wavenumber regions that allow surface modes. A positive value of real  $\beta $ is needed for the solution of surface modes in Eq. (\ref{solb}). Using the above parameters, the bulk mode region is bounded by  $k=0$ cm\textsuperscript{{}-1} and \textit{k=}16\textit{} cm\textsuperscript{{}-1} as shown in Fig. \ref{attenuation}, and is indicated in Fig. \ref{dispersion} by shading. Comparison of Figs. \ref{attenuation} and \ref{dispersion} shows that the resonance at $\omega _{m}$${\cong}$ 3 cm\textsuperscript{{}-1} is associated with a divergence of the attenuation constant $\beta $ due to the zero of  $\mu _{z}$. From this zero condition, an expression for  $\omega _{m}$ can be derived :
\begin{equation}
	\omega _{m}=\left[\left(\omega _{\mathit{afm}}^{2}\cos ^{2}\theta
+\Omega _{\mathit{me}}^{2}\right)+\left(2\omega _{s}\omega
_{\mathit{me}}\sin 2\theta
\right)\right]^{1/2}
\end{equation}
with  $\omega _{s}=\gamma 4\pi M_{s}$. 

The pole and zero frequency of bulk modes,  $\omega _{p}$ and  $\omega
_{z}$, \ is related to the zeroes of  $\mu _{y}$ and \ \  $f(\mu )=\mu
_{y}\mu _{z}-\mu _{\mathit{yz}}^{2}$ . These give
\begin{equation}
	\omega _{p}=\left[\left(\omega _{\mathit{afm}}^{2}\cos ^{2}\theta
+\Omega _{\mathit{me}}^{2}\right)+\left(2\omega _{s}\omega _{a}\cos
2\theta \right)\right]^{1/2}
\end{equation}
and
\begin{equation}
	\omega _{z}=\frac{1}{\sqrt{2}}\left[\Omega
_{\mathit{mp}}^{2}+\left(\Omega _{\mathit{mp}}^{4}-4\omega
_{m}^{2}\omega
_{p}^{2}\right)^{1/2}\right]^{1/2}
\end{equation}
where
\begin{equation}
	\Omega _{\mathit{mp}}^{2}=\omega _{m}^{2}+\omega _{p}^{2}+4\omega
_{s}^{2}\sin ^{2}\theta 
\end{equation}
These three frequencies, $\omega _{m}$, $\omega _{p}$ and  $\omega_{z}$ divide the bulk region into the shaded regions shown in Fig.\ref{dispersion}. Note that the lowest two bands overlap, and are deformed
strongly near the resonance $\omega _{m}$.

Surface modes, which is indicated by {\textquotedblleft}SP{\textquotedblright} in Fig. \ref{dispersion} can exist in the gaps between  $\omega _{m}$ and  $\omega_{p}$.  The surface branches are nonreciprocal with respect to propagation, with  $\omega\left(\vec{k}\right)\neq \omega (-\vec{k})$. The negative branch begins from the intersection of the lightline with the  $\omega _{m}$ resonance, and is indicated by {\textquotedblleft}-L{\textquotedblright} . The branch terminates in the left middle bulk band. The positive branch begins at the lightline with frequency
\begin{equation}
	\omega =\left[\frac{\epsilon _{x}\omega _{m}^{2}-\omega
_{\mathit{afm}}^{2}\cos ^{2}\theta +\Omega _{\mathit{me}}^{2}}{\epsilon
_{x}-1}\right]^{1/2}
\end{equation}
which is slightly higher than $\omega _{m}$, and then terminates at the upper bulk band.

A method for detecting surface modes at THz frequencies is Attenuated Total Reflection (ATR)\cite{camley82} and has been used to study surface modes on antiferromagnets\cite{jensen95}. In this method, a high optical index prism is used to couple electromagnetic radiation to surface excitations that lie off the vacuum light line.  A sharp dip of reflectivity indicates excitation of surface modes. Using results from Ref. [\onlinecite{abraha96}], reflectivity in ATR is given by
\begin{equation}
	R=\left|\frac{k_{z}\left(1+re^{-2\beta _{o}d}\right)-i\beta
_{o}\left(1-re^{-2\beta _{o}d}\right)}{k_{z}\left(1+re^{-2\beta
_{o}d}\right)+i\beta _{o}\left(1-re^{-2\beta _{o}d}\right)}\right|^{2}
\end{equation}
where \textit{d} represents the distance between prism and the sample, and \textit{r} is defined as
\begin{equation}
	r=\frac{\beta _{o}-\kappa }{\beta _{o}+\kappa}
\end{equation}
with  
\begin{equation}
	\kappa =\frac{\mu _{z}\beta -\mu_{\mathit{yz}}k_{y}}{\mu _{y}\mu _{z}-\mu_{\mathit{yz}}^{2}}.
\end{equation}
Here, the wave-vector
\begin{equation}
	k_y=(\epsilon_p)^{1/2}\frac{\omega}{c}\sin\theta
\end{equation}
and
\begin{equation}
	k_{z}=(\epsilon_{p})^{1/2}\frac{\omega}{c}\cos\theta
\end{equation}
represents propagation along and normal to the surface, where $\epsilon _{p}$ is the prism dielectric constant, and  $\theta $ is the incident angle.

%--------------------------------------------------------------------------------------------------------figure 3 --------------------------%
\begin{figure}[ht]
\begin{center}
\subfigure[\label{30}ATR spectra at 30$^o$.]{
\includegraphics[width=7cm]{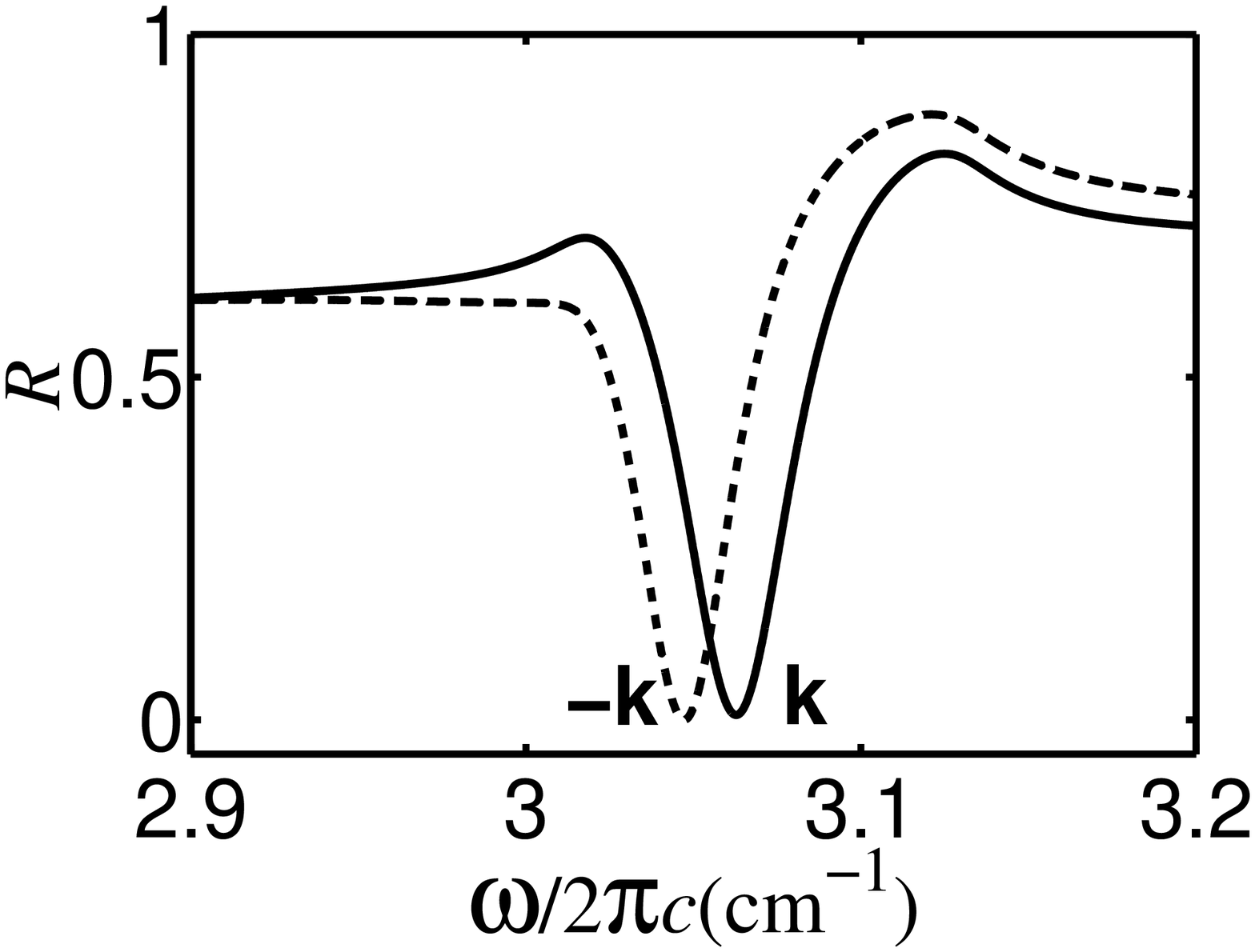}}
\subfigure[\label{70}ATR spectra at 70$^o$.]{
\includegraphics[width=7cm]{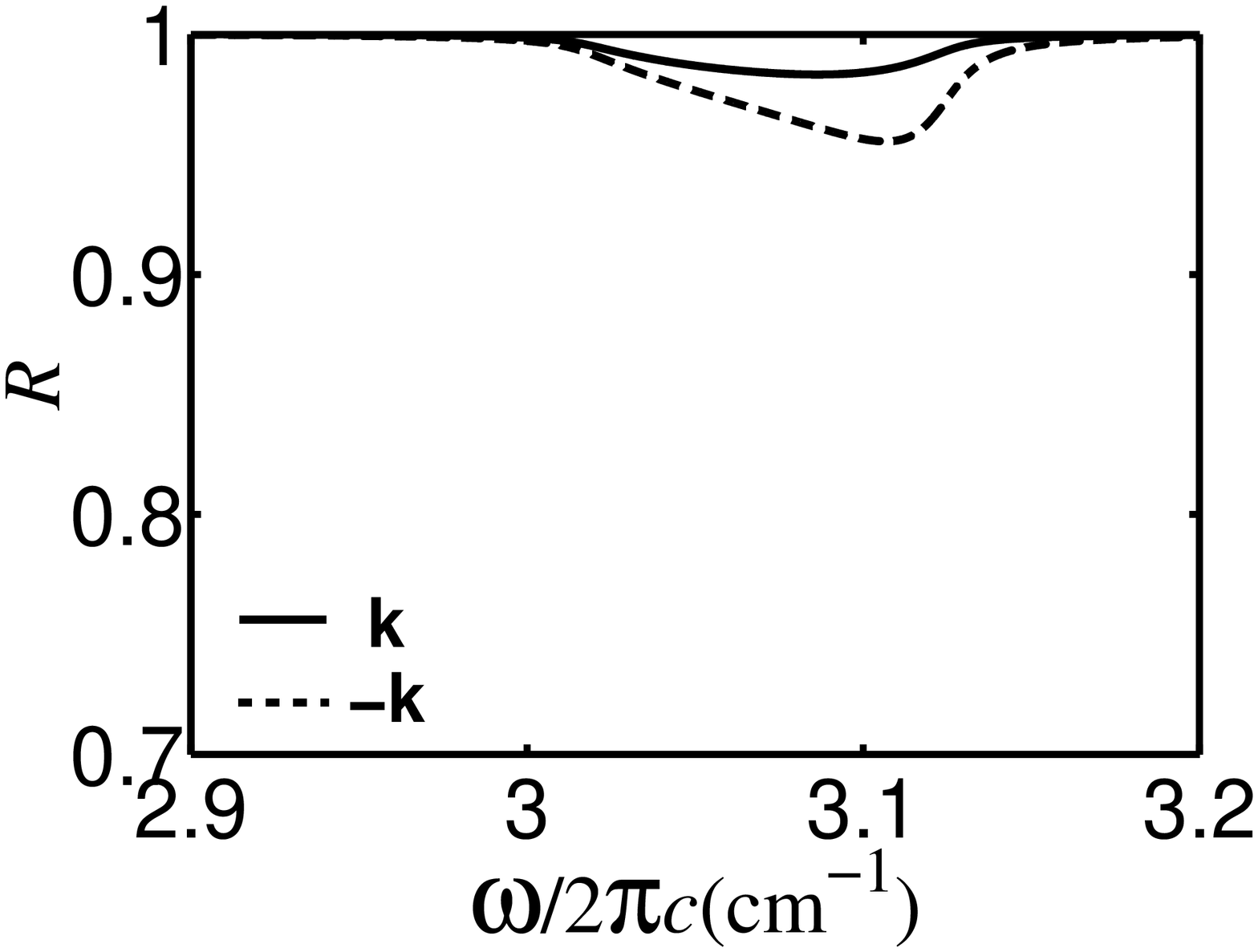}}
\caption{ATR spectra with incident angles  30$^o$ and 70$^o$. In (a) two different sharp dips illustrate the non-reciprocity of the surface modes. In (b) the absence of the sharp dips indicate the absence of surface modes.}
\end{center}
\end{figure}
%----------------------------------------------------------------------------------------------------------------------------------%

Calculated ATR results for the BaMnF$_4$ is presented in Fig. \ref{30} and \ref{70} for the incident angles 30$^o$ and 70$^o$. The ATR spectra at 30$^o$ illustrates nicely the nonreciprocity of surface modes. Two sharp dips correspond to surface polaritons traveling in the opposite directions and the frequencies are the  $\omega_{m}$ intersection points discussed above for the positive and negative surface mode branches. The ATR for an incident angle of 70$^o$, shown in Fig.\ref{70} does not allow coupling to surface modes.  The shallow dips represent bulk modes.

It is interesting to notice that the surface modes do not exist in the region where the wave-vector   $k\gg \frac{\omega }{c}$. In this region, the surface dispersion relation in Eq. (\ref{surface}) can be re-written in the form
\begin{equation}
\label{mgnstat}
\mu _{y}\mu _{z}-\left(\mu _{\mathit{yz}}\mp1\right)^{2}=0
\end{equation}
for $k=\pm\infty $ . In terms of $\omega _{m}$,  $\omega_{p}$ and  $\omega _{r}$ Eq. (\ref{mgnstat}) becomes
\begin{equation}
\label{mgnstat2}
\left(\omega _{m}^{2}-\omega ^{2}\right)\left(\omega
_{p}^{2}-\omega ^{2}\right)-\left[2\omega _{s}\omega \sin \theta \mp
\left(\omega _{r}^{2}-\omega ^{2}\right)\right]^{2}=0
\end{equation}
where $\omega _{r}=\left(\omega _{\mathit{afm}}^{2}\cos ^{2}\theta+\Omega_{\mathit{me}}^{2}\right)^{1/2}$ represents the pole frequency of permeabilities given by Eq.(\ref{dyn1}-\ref{dyn2}) and is located slightly below  $\omega _{m}$ so that  $\omega _{r}<\omega _{m}<\omega _{p}.$ Assume that the asymptotic
frequency for the surface is located in the region above the pole frequency  $\omega _{p}$. For $k=+\infty $, the second term in Eq. (\ref{mgnstat2}) is always larger than the first term. Hence, Eq. (\ref{mgnstat2}) is never satisfied and the surface modes do not exist in this wavelength region.
%--------------------------------------------------------------------------------------------------------figure 4 --------------------------%
\begin{figure}[ht]
\begin{center}
\subfigure[\label{nomgnt}The absence of surface modes at $-k$.]{
\includegraphics[width=6.5cm]{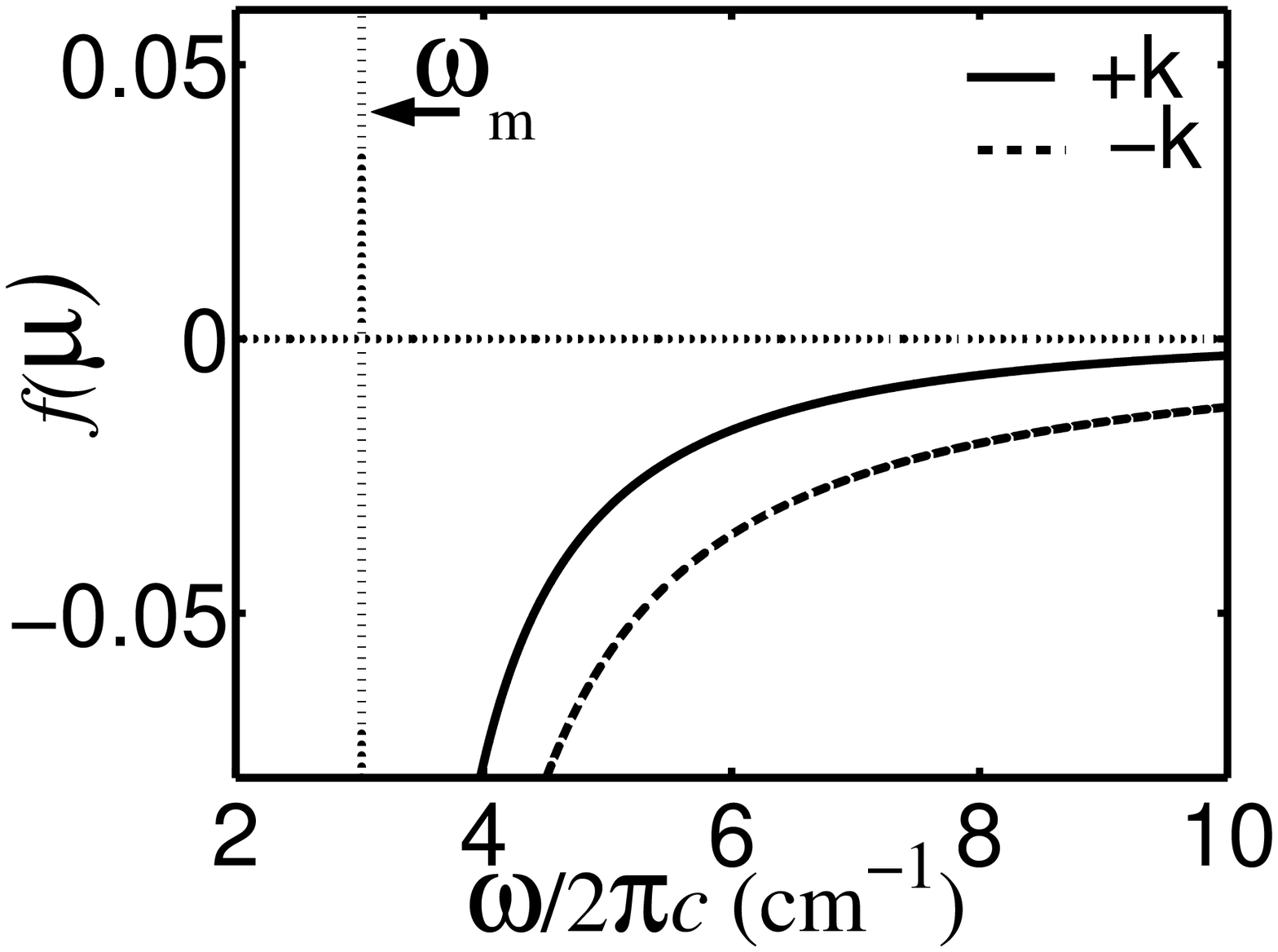}}
\subfigure[\label{yesmgnt}The existence of surface modes at $-k$.]{
\includegraphics[width=6.25cm]{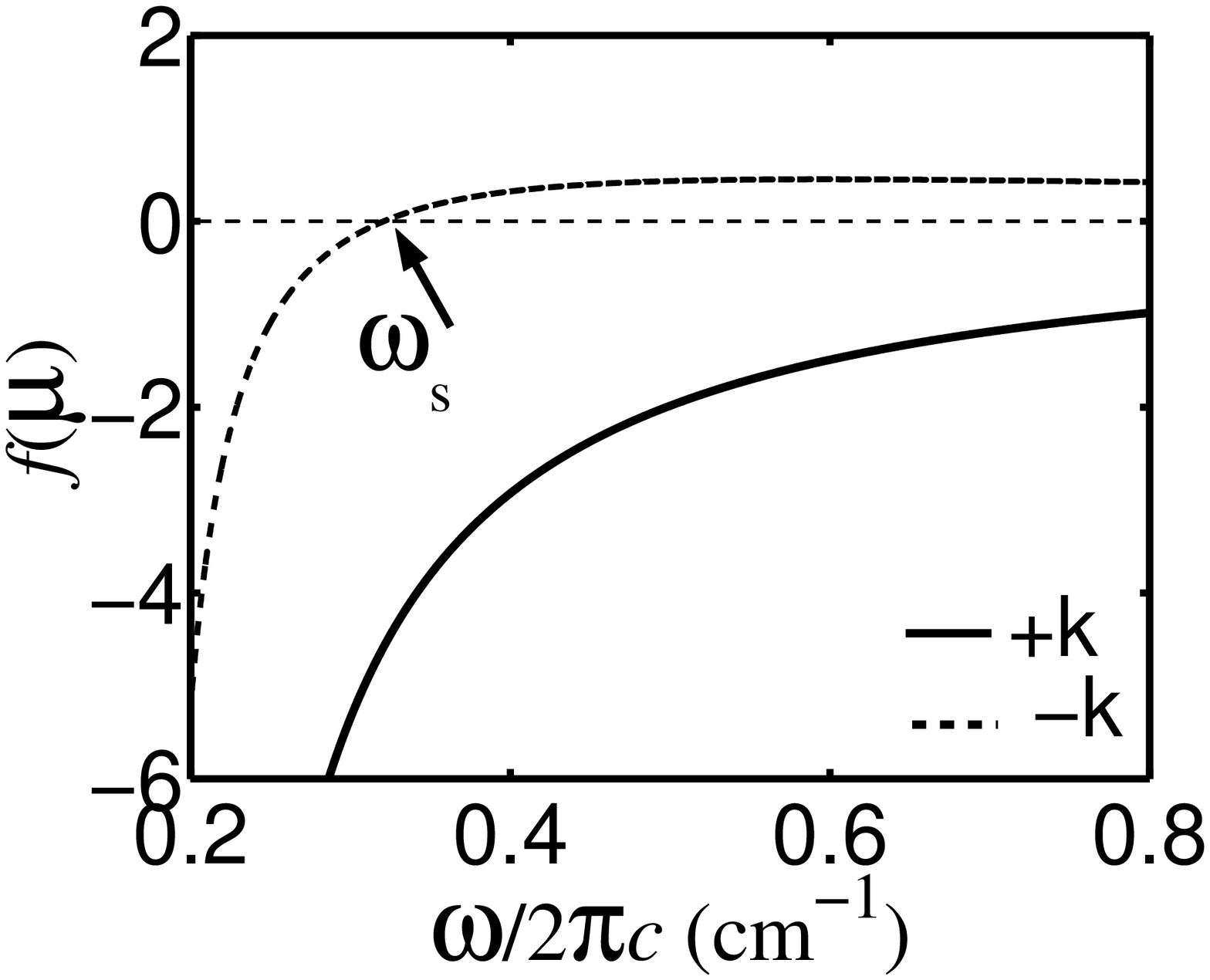}}
\subfigure[\label{15}The parameters are $\lambda/\alpha$=1.5.]{
\includegraphics[width=6.25cm]{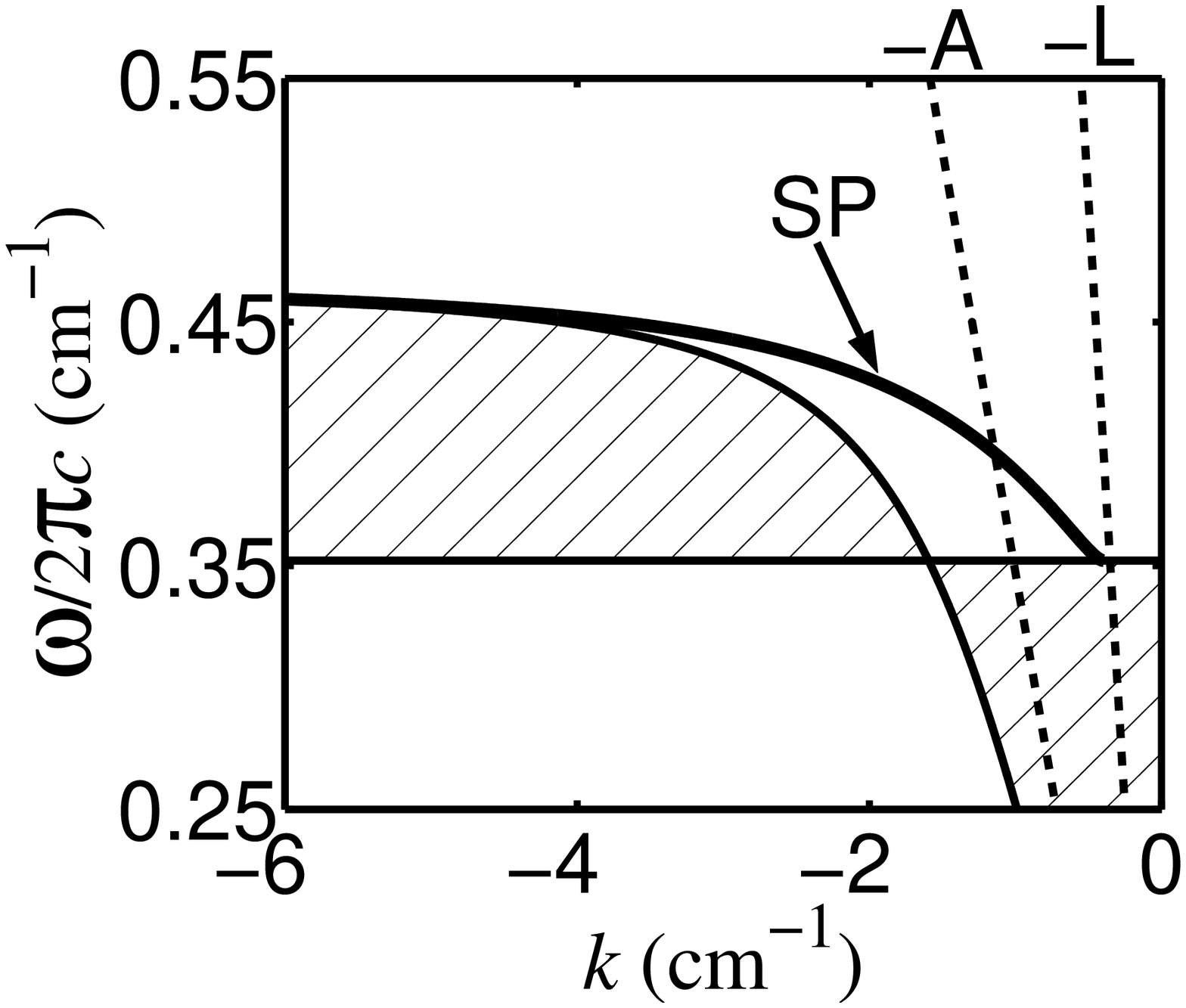}}
\subfigure[The parameters are $\lambda/\alpha$=2.]{
\includegraphics[width=6.25cm]{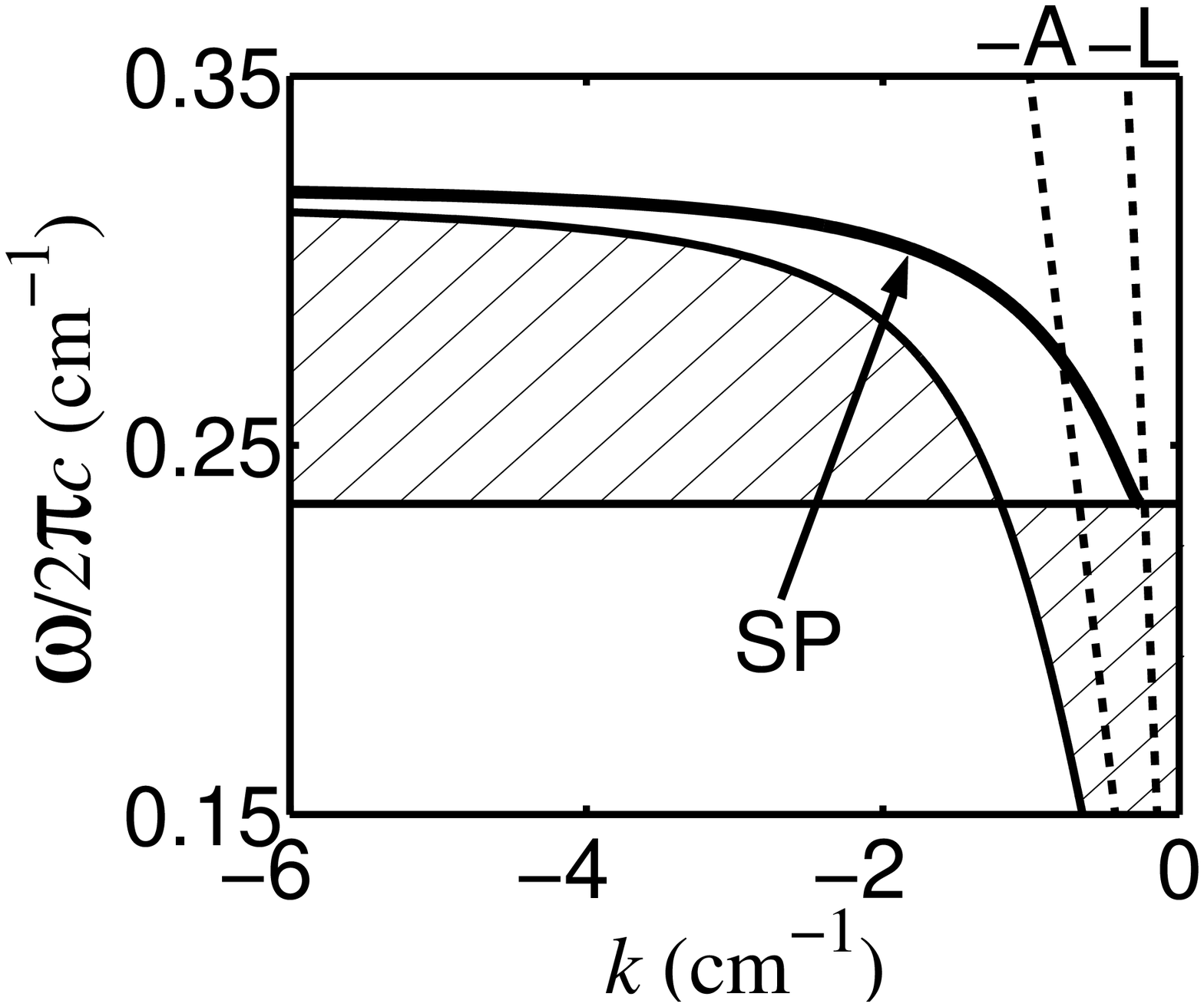}}
\caption{\label{mgnt}  The function in Eq. (\ref{mgnstat}) represented by $f(\mu)$ is shown in (a) using parameters for BaMnF$_4$. The   condition for  surface modes in $k>>\omega/c$ region has no  solution.  In (b) a solution exists using parameters where $\lambda/\alpha=1.5$, where the surface begin to overcome the bulk as shown in (c). In (d) the surface branch is slightly higher than bulk when the ratio increase to $\lambda/\alpha=2$.}
\end{center}
\end{figure}
%----------------------------------------------------------------------------------------------------------------------------------%

A surface mode at  $k=-\infty$ may exist for some values of  $\omega_{r},\omega _{\mathit{m}}$ and  $\omega _{p}$.  For a solution to exist in this region, the function  $f(\mu )$ of Eq. (\ref{mgnstat2}) should cross zero. This gives a condition from Eq. (\ref{mgnstat2}) at the frequency $\omega =\omega _{p}$ for a surface mode solution to exist. The requirement that $f\left(\mu\right)$ vanish means that 
\begin{equation}
\label{syarat}
\omega _{p}\le \omega _{a}\cos 2\theta +2\omega_{\mathit{me}}\sin 2\theta
\end{equation}

For BaMnF$_4$, the requirement in Eq. (\ref{mgnstat}) is never satisfied as shown in Fig. \ref{nomgnt}, hence there is no surface modes at $k>>\frac{\omega}{c}$. The condition Eq. (\ref{syarat}) can be satisfied by changing, for example, the parameters so that  $\lambda/\alpha=1.5$.  An example is shown in Fig. \ref{yesmgnt} , and the associated dispersion relation is presented in Fig.\ref{15}. In this case a surface mode is able to exist at the top of the middle bulk band, for arbitrarily small wavelengths.

%--------------------------------------------------------------------------------------------------------figure 5 --------------------------%
\begin{figure}[ht]
\begin{center}
\subfigure[\label{sdtH}canting angle versus $H$]{
\includegraphics[width=6.5cm]{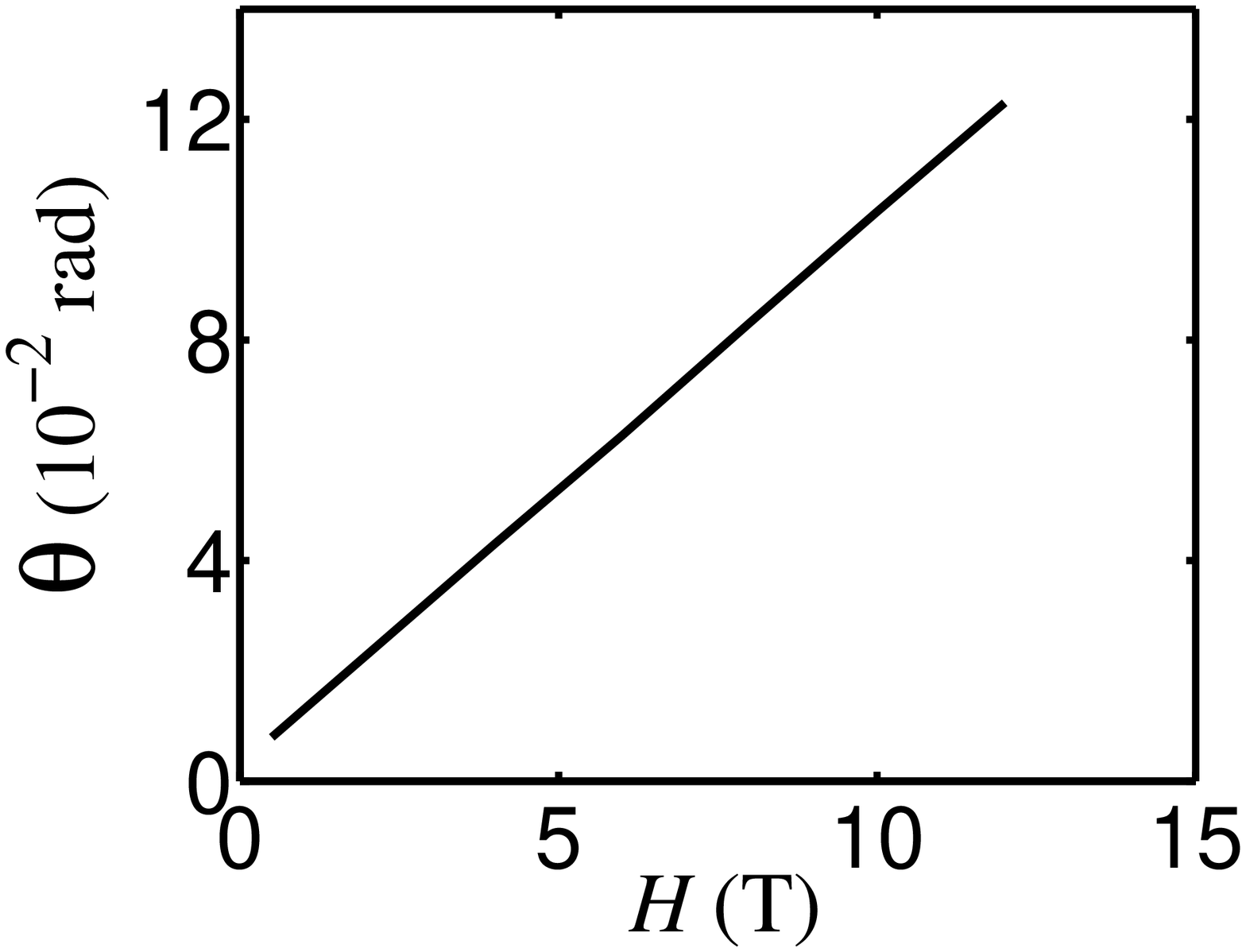}}
\subfigure[\label{polH}Surface modes at $H=15$T]{
\includegraphics[width=7cm]{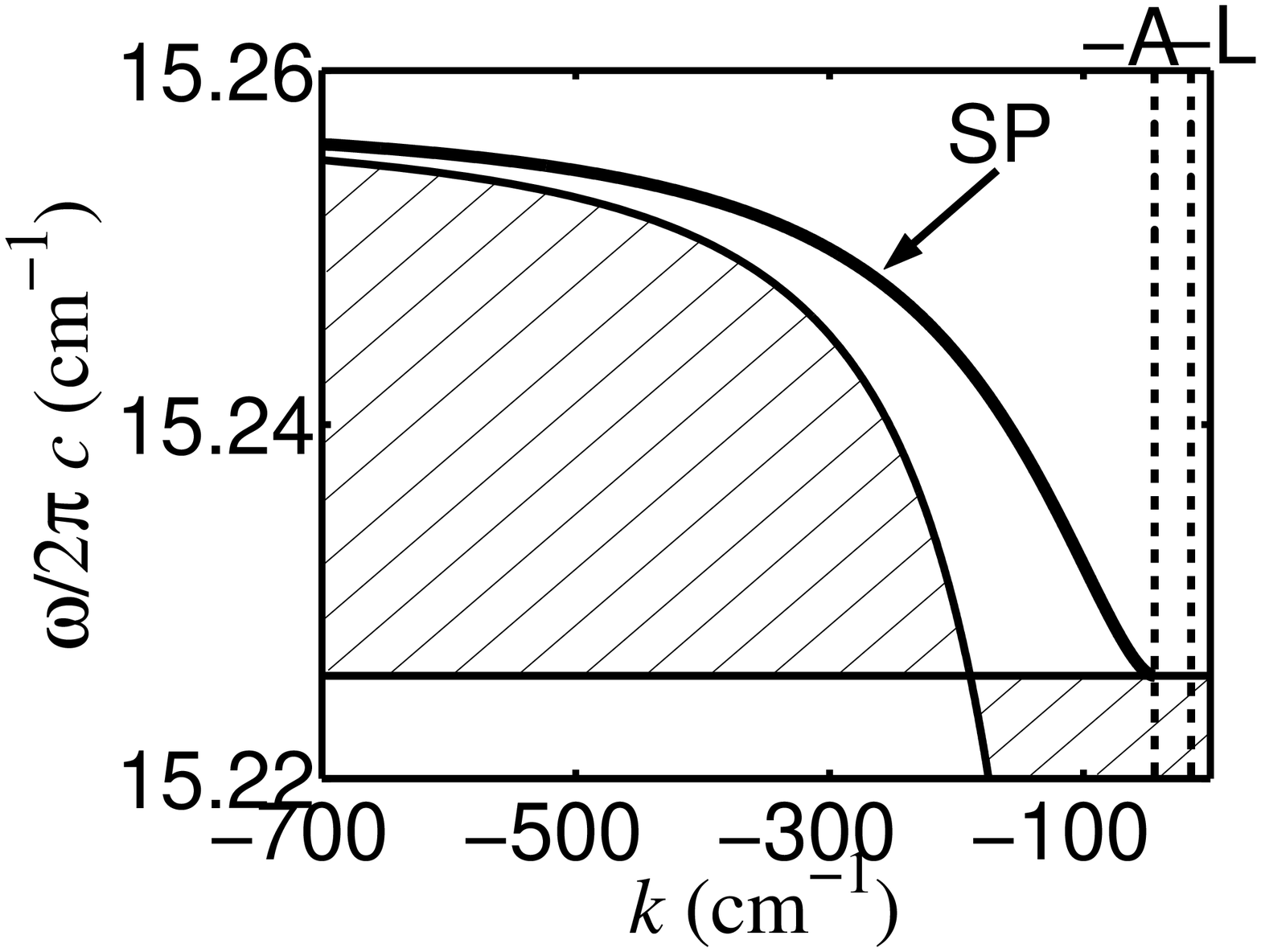}}
\caption{The influence of magnetic field.  In (a)influence on  canting angle.  In (b), the application of a magnetic external field of 15 T increases and allows the negative surface branch to enter the magnetostatic region, where $k>>\frac{\omega}{c}$.}
\end{center}
\end{figure}
%----------------------------------------------------------------------------------------------------------------------------------%

In principle, application of an external field can be used to modify the surface mode frequencies by changing the canting angle. However, as mention above that the electric part is uncoupled from the magnetic part, and application of an electric field has negligible effect on the polariton frequencies. 

The application of an external static magnetic field changes directly the canting angle and thereby affect the frequencies. The canting angle as a function of applied magnetic field shown in Fig. \ref{sdtH}. In fact, for sufficiently strong magnetic field, the condition of Eq. (\ref{mgnstat}) can be satisfied and a surface modes branch extend to  $k=-\infty $.  This occurs through an increase in $\omega _{r}$ via the field dependence of  $\Omega _{o}$(see Eq. (\ref{Wh})).

The frequencies  $\omega _{m}$ and pole frequency  $\omega _{p}$ also increase under an external magnetic field. The difference between $\omega _{m}$ and  $\omega _{p}$ is given approximately by
\begin{equation}
	\Delta ^{2}\approx \left(\omega _{p}^{2}-\omega
_{m}^{2}\right)\propto \left(\cos 2\theta -\sin 2\theta
\right)
\end{equation}
As a consequence, the middle bulk band narrows in frequency with increasing magnetic field. At a field of 15 T, the BaMnF$_4$ middle bulk band is sufficiently narrow that a surface mode rises above the middle band and extends to small wavelengths, as illustrated in Fig. \ref{polH}.

Lastly, we consider the case where the dielectric constant depends on frequency. We suppose that the frequency dependence is given by the soft phonon mode according to
\begin{equation}
	\omega _{T}^{2}=\left(\beta _{1}+P^{2}_{o}\beta
_{2}\right)f
\end{equation}
with the dielectric function
\begin{equation}
\epsilon _{x}=\epsilon _{b}\frac{\omega _{L}^{2}-\omega
^{2}}{\omega _{T}^{2}-\omega
^{2}}
\end{equation}
Here  $\omega _{L}^{2}=\omega _{T}^{2}+\frac{f}{\epsilon _{o}\epsilon_{b}}$ and  $\epsilon _{b}=8.3$ is dielectric constant background. 

%--------------------------------------------------------------------------------------------------------figure 6 --------------------------%
\begin{figure}[ht]
\begin{center}
\subfigure[\label{ehigh}The case where $\omega_{e}>\omega_{m}$.]{
\includegraphics[width=7cm]{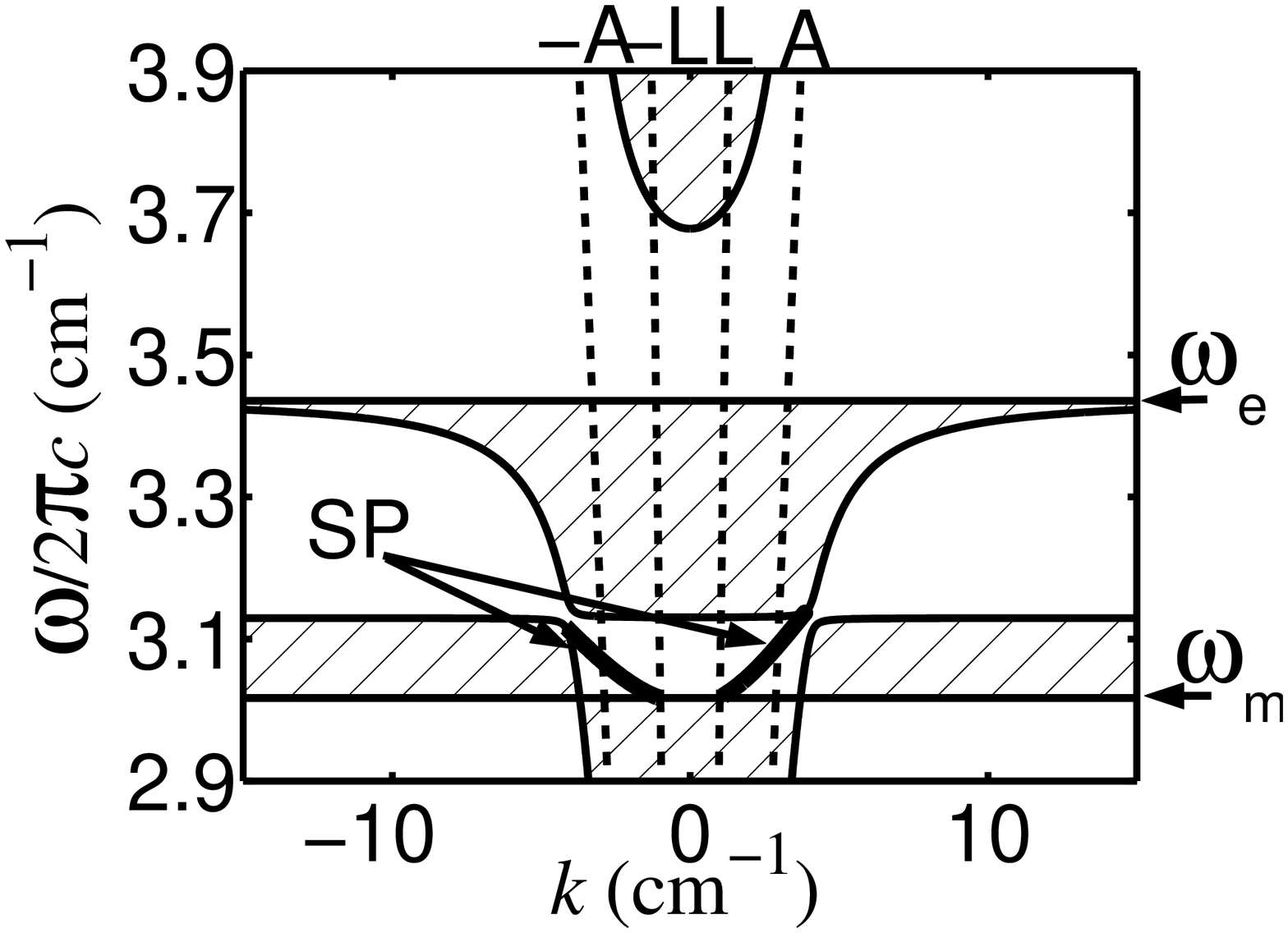}}
\subfigure[\label{mhigh}The case where $\omega_{m}>\omega_{e}$.]{
\includegraphics[width=7cm]{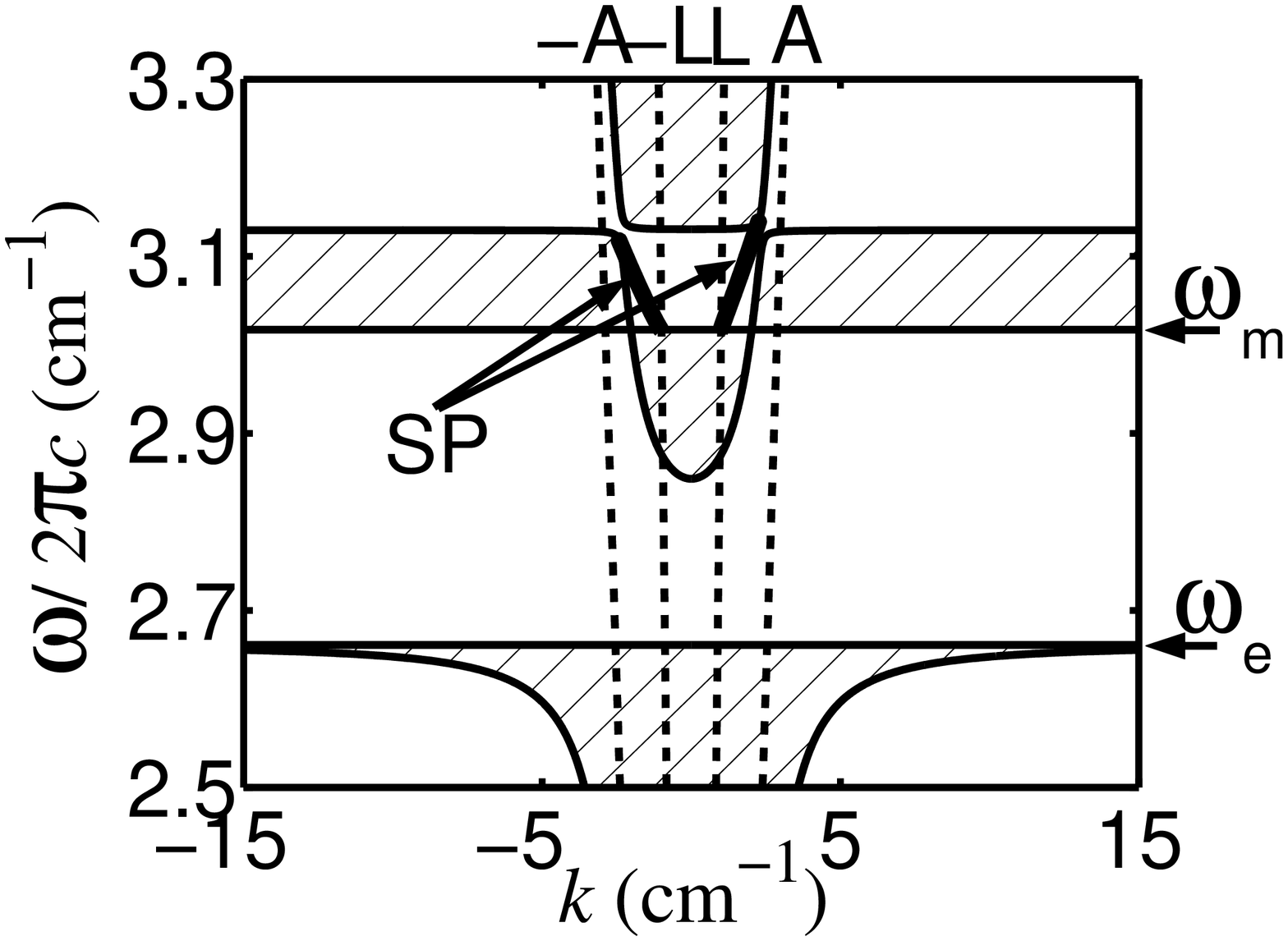}}
\caption{Dispersion relation for a frequency dependent dielectric function.  The frequency is introduced according to  Eq. (35).   In (a), the case when of a magnetic resonance frequency lower than the electric resonance frequency is shown.  In (b), the case of a magnetic resonance frequency  higher than the electric resonance frequency is shown.}
\end{center}
\end{figure}
%----------------------------------------------------------------------------------------------------------------------------------%

Since there is no dynamic electric and magnetic coupling for the TE modes, there is no effect on the surface mode frequencies even though the electric resonance is near the magnetic resonance in frequency
(this can be achieved by modifying the phonon mass). Effects on the bulk bands for an inverse of the phonon mass of 0.015 \textit{f } is shown in Fig. \ref{ehigh}, where the magnetic resonance is lower than the dielectric resonance. The effects of a magnetic resonance above the dielectric resonance, with 0.009\textit{f, }are shown in Fig. \ref{mhigh}.  Only band overlap and intersections are affected. Note a consequence of this for the surface modes. In the case where the magnetic resonance is below the dielectric resonance, have a broader range in wavevector (see Fig. \ref{ehigh}) compared to the case where the magnetic resonance is above the dielectric resonance (in Fig. \ref{mhigh}).

%----------------------------------------------------------------------------------------------------------------------------------%
 
\section{Conclusions}
\label{Conclusions} 

Dispersion relations for a material having linear magnetoelectric coupling have been calculated for tranverse electric polariton modes. Effects associated with spin canting are considered. The polarisation and weak ferromagnetism are parallel to the surface, and the antiferromagnet magnetisations are out of plane. We find surface modes associated with the weak ferromagnetism that are non-reciprocal with respect to propagation direction, such that $\omega \left(\vec{k}\right)\neq\omega (-\vec{k})$. For sufficiently large magnetic fields, or different material parameters, surface modes may also exist in the magnetostatic region, with  $k\gg \omega /c$. Application of static magnetic field can be used to modify the middle bulk band frequencies

\begin{acknowledgments}
We wish to acknowledge the support of Ausaid, the Australian Research Council and DEST.
\end{acknowledgments}

%Merlin.mbs v4.21 2009-07-09.
%
%======================%
\end{document}